\documentclass[aip,jcp,reprint]{revtex4-1}

\usepackage{graphicx}
\usepackage{amsmath}
\usepackage{latexsym}
\usepackage{hyperref}

\begin{document}

\noindent
Published as J. Chem. Phys. {\bf 139}, 164701 (2013). 

\noindent
See \url{http://dx.doi.org/10.1063/1.4825176}.

\title{Local order variations in confined hard-sphere fluids}

\author{Kim~Nyg{\aa}rd}
\email[]{kim.nygard@chem.gu.se}
\affiliation{Department of Chemistry and Molecular Biology, University of Gothenburg, 
SE-412 96 Gothenburg, Sweden}

\author{Sten~Sarman}
\email[]{sarman@ownit.nu}
\affiliation{Department of Materials and Environmental Chemistry, 
Stockholm University, SE-106 91 Stockholm, Sweden}

\author{Roland~Kjellander}
\email[]{rkj@chem.gu.se}
\affiliation{Department of Chemistry and Molecular Biology, University of Gothenburg, 
SE-412 96 Gothenburg, Sweden}

\date{\today}

\begin{abstract}
Pair distributions of fluids confined between two surfaces at close distance are of fundamental 
importance for a variety of physical, chemical, and biological phenomena, such as interactions 
between macromolecules in solution, surface forces, and diffusion in narrow pores. However, 
in contrast to bulk fluids, properties of inhomogeneous fluids are seldom studied at the 
pair-distribution level. Motivated by recent experimental advances in determining anisotropic 
structure factors of confined fluids, we analyze theoretically the underlying anisotropic pair 
distributions of the archetypical hard-sphere fluid confined between two parallel hard surfaces 
using first-principles statistical mechanics of inhomogeneous fluids. 
For this purpose, we introduce an experimentally accessible ensemble-averaged local density correlation function and study its behavior as a function of confining slit width. 
Upon increasing the 
distance between the confining surfaces, we observe an alternating sequence of strongly 
anisotropic versus more isotropic local order. The latter is due to packing frustration of the 
spherical particles. This observation highlights the importance of studying inhomogeneous 
fluids at the pair-distribution level. 
\end{abstract}

\maketitle

\section{Introduction}

Fluids confined between two surfaces at close distance are abundant in physical, chemical, 
and biological systems. The spatial confinement induces a complex microscopic ordering of 
the fluid, which depends on both the interactions between the fluid particles and the confining 
walls as well as the mutual interactions between the fluid particles. The microscopic  
structure of the fluid is of fundamental importance for a wide range of phenomena, such 
as the interactions between macromolecules or colloidal particles in solution,\cite{israelachvili96}  
apparent charge reversal of suspended particles due to many-body ion 
correlations,\cite{torrie82,kekicheff93,kjellander09,kubickova12} and 
like-charge attraction due to ion-ion correlations.\cite{guldbrand84,kjellander84a,kjellander88c} 
Moreover, an accurate description of the dynamical properties 
of confined fluids, such as the diffusivity in narrow pores\cite{mittal08,lang10} and the friction 
between surfaces suspended in fluids,\cite{bhushan95} necessitates a good description of 
static structure of the fluid. The relevance of the topic is further highlighted by the development 
of novel technological applications based on confinement of fluids, such as 
ionogels.\cite{neouze06}  

Fluids are disordered systems, which are characterized by short-range density variations 
known as the local structure of the fluid. In the case of bulk fluids of spherical particles, 
the isotropic density around a fluid particle is given by $n_0 g(r)$, with $n_0$ denoting the 
bulk number density, $g(r)$ the pair-distribution function (also called the radial distribution function), 
and $r$ the distance from the particle 
center. The pair distributions are also directly related to thermodynamic quantities, thereby 
providing a formal connection between microscopic and macroscopic properties of the 
fluid.\cite{hansen06} Moreover, pair distributions can be routinely determined experimentally 
by means of x-ray or neutron scattering, making them the single most important quantity for 
characterization of fluid properties.   

For confined fluids, in turn, the local density is governed by a complex interplay between 
particle-wall and particle-particle interactions. Hence, the local density at position ${\bf r}_1$ 
around a particle with its center at position ${\bf r}_2$ is given by $n({\bf r}_1)g({\bf r}_1,{\bf r}_2)$, 
with $n({\bf r}_1)$ denoting the number density profile and $g({\bf r}_1,{\bf r}_2)$ 
the pair-distribution function. In comparison with bulk fluids, there are two notable differences 
due to the presence of the confining surfaces: (i) $n({\bf r}_1)$ exhibits spatial variation and 
(ii) $g({\bf r}_1,{\bf r}_2)$ 
is anisotropic, depending on the individual values of ${\bf r}_1$ and ${\bf r}_2$ rather than on 
the magnitude of their difference, $r_{12} = \vert {\bf r}_1 - {\bf r}_2 \vert$. Moreover, while the 
properties of bulk fluids are routinely analyzed microscopically in terms of their pair distributions, 
studies on the pair distributions of confined fluids remain scarce. 

Given their fundamental importance, one may question why the pair distributions of confined 
fluids have, to a large extent, been neglected so far? This neglect can primarily be attributed to 
two causes. First, although the theoretical framework was developed a long time 
ago,\cite{percus64,sokolowski80} the determination of theoretical pair distributions in 
confined fluids has so far been considered a computationally demanding task. In fact, 
the vast majority of computational work has focused on the simpler and generally 
less accurate singlet distribution (i.e., the density profile), while explicit calculations of pair 
distributions in confined fluids have been reported only very 
seldom.\cite{kjellander88a,kjellander91,kjellander91b,gotzelmann97,henderson97, botan09,zwanikken13}  
Second, there has to date been a lack of experimental data for comparison at the 
pair-distribution level, and hence there has been no strong incentive to explicitly determine 
theoretical pair distributions of confined fluids. Experimental studies have instead focused on  
singlet distributions of confined fluids - either indirectly using surface-force 
experiments\cite{horn81,israelachvili87,klapp08} or directly using, e.g., 
x-ray scattering\cite{zwanenburg00} or confocal microscopy.\cite{nugent07}  
However, recently we demonstrated an experimental approach based on x-ray scattering 
from colloid-filled nanofluidic channel arrays,\cite{nygard09a} providing experimental 
access to confined fluids at the pair-distribution level in terms of anisotropic structure factors 
- in quantitative agreement with first-principles statistical mechanics of inhomogeneous 
fluids.\cite{nygard12} 

In this paper, we analyze theoretically the anisotropic structure factors of Ref.~\onlinecite{nygard12} 
in terms of the underlying pair distributions. As a model system we study the archetypical 
hard-sphere fluid confined between smooth and hard planar surfaces, which is a good 
approximation for entropy-dominated fluids. The first-principles theoretical calculations are 
carried out within integral-equation theory, by solving the inhomogeneous Ornstein-Zernike 
and Lovett-Mou-Buff-Wertheim equations using the anisotropic Percus-Yevick 
closure.\cite{kjellander91,kjellander88b} The main results of the paper are two-fold:
First, we show that the experimentally accessible anisotropic structure factor can be 
interpreted in terms of an ensemble-averaged local density correlation function of the confined fluid. 
Next, we use this result to interpret the experimental findings of Ref.~\onlinecite{nygard12} as evidence 
for an alternating sequence of highly anisotropic, periodically modulated versus a more isotropic 
local order upon increasing the separation between the confining surfaces. 
In essence, this effect is driven by packing frustration, i.e., an incompability between the 
preferred local order of the fluid and the layering induced by the confining surfaces. The 
direct observation of a hitherto unknown sequence of local ordering-disordering phenomena 
on the pair-distribution level in the extensively studied system of a hard-sphere fluid between 
hard planar surfaces emphasizes the importance of explicitly studying inhomogeneous fluids 
at this level.

\section{Methods}

\subsection{Inhomogeneous integral-equation theory}

The interaction potentials for the system with hard spheres between two hard surfaces 
are given by the particle-particle interaction potential, 
\begin{equation} 
\beta u({\bf r}_1,{\bf r}_2) = \left\{ \begin{array}{ll} 
0 & \textrm{if $\vert {\bf r}_1 - {\bf r}_2 \vert \geq \sigma$}\\ 
\infty & \textrm{if $ \vert {\bf r}_1 - {\bf r}_2 \vert < \sigma$}
\end{array} \right.
\label{eq:particle}
\end{equation} 
and the particle-wall potential, 
\begin{equation} 
\beta v(z) = \left\{ \begin{array}{ll} 
0 & \textrm{if $\vert z \vert \leq L/2$,}\\ 
\infty & \textrm{if $ \vert z§ \vert > L/2$,}
\end{array} \right. 
\label{eq:particle_wall}
\end{equation} 
with $\beta = (k_B T)^{-1}$, $k_B$ Boltzmann's constant, and $T$ the absolute temperature. 
Here, the $z$-axis is placed perpendicular to the confining surfaces with its origin midway 
in between, while the particle centers are confined to a reduced slit width of $L = H-\sigma$, 
with $H$ denoting the surface separation and $\sigma$ the particle diameter. 

In the calculations  the planar symmetry of the system has been utilized. This reduces the spatial 
dimension of the density distribution $n({\bf r})$ from three to one and the pair-distribution 
functions from six to three, for 
example $g({\bf r}_1,{\bf r}_2)=g(z_1,z_2,R_{12})$, where $R_{12}=|{\bf R}_{12}|$ and
${\bf R}_{12}=(x_2-x_1,y_2-y_1)$. The density profile, the total correlation function 
$h({\bf r}_1,{\bf r}_2) = h(z_1,z_2,R_{12}) = g(z_1,z_2,R_{12})-1$, and the 
direct correlation function $c({\bf r}_1,{\bf r}_2)=c(z_1,z_2,R_{12})$  are obtained by solving 
the following set of equations: the Lovett-Mou-Buff-Wertheim equation, 
\begin{equation}
\frac{d [\log n(z_1)+\beta v(z_1)]}{dz_1}=
 \int c(z_1,z_2,R_{12})\frac{dn(z_2)}{dz_2} dz_2 d{\bf R}_{12},  
\label{eq:LMBW}
\end{equation} 
and the inhomogeneous Ornstein-Zernike equation, 
\begin{equation}
h({\bf r}_1,{\bf r}_2)=c({\bf r}_1,{\bf r}_2)+
\int h({\bf r}_1,{\bf r}_3)n(z_3)c({\bf r}_3,{\bf r}_2) d{\bf r}_3, 
\label{eq:OZ}
\end{equation}
subject to the anisotropic Percus-Yevick (PY) closure
\begin{equation}
c({\bf r}_1,{\bf r}_2) = g({\bf r}_1,{\bf r}_2) - y({\bf r}_1,{\bf r}_2), 
\label{eq:PY}
\end{equation}
where $y({\bf r}_1,{\bf r}_2)$ is the cavity function that satisfies
\begin{equation}
g({\bf r}_1,{\bf r}_2) = y({\bf r}_1,{\bf r}_2) \exp [-\beta u({\bf r}_1,{\bf r}_2)]. 
\label{eq:cavity}
\end{equation}
The PY closure is the only approximation made. The set of Eqs.~(\ref{eq:LMBW}) - (\ref{eq:cavity})
is solved fully self-consistently in an iterative manner.

In the calculations  the cavity function $y$ is determined numerically on a grid. The number of 
grid points can thereby be kept to a minimum, since this function is continuous at the hard 
core periphery of the spheres. The pair-distribution function is then obtained from 
Eq.~(\ref{eq:cavity}).

\subsection{Boundary conditions}

To solve Eq.~(\ref{eq:LMBW}), one needs a 
boundary condition for the density profile or some other suitable information. There are two 
particularly convenient choices: the number of particles per unit area in the slit, 
$N=\int_{-L/2}^{L/2}n(z)dz$, or the value of the density at some point, for instance the 
contact density at a wall surface $n(\pm L/2)$ or the value at the midpoint between the 
surfaces $n(0)$.  One must, however, know what value to use when the fluid in the slit is in 
equilibrium with a bulk fluid of given density, which is a nontrivial problem 
within most integral equation theories at the anisotropic pair distribution level.\cite{note_PY} We used the 
following method to determine this value for various surface separations. 

The rate of change of the profile when the surface separation is changed under the condition of constant chemical potential  is given by the exact equation\cite{kjellander88b, kjellander91}
\begin{multline}
\frac{\partial n(z_1;L)}{\partial L}=-\beta n(z_1;L) \bigg[ \frac{\partial v(z_1;L)}{\partial L}+ 
 \int n(z_2;L)  \\
 \times h(z_1,z_2,R_{12};L)\frac{\partial v(z_2;L)}{\partial L} dz_2 d{\bf R}_{12} \bigg],  
\label{eq:profile_equil}
\end{multline}
where we have indicated explicitly that all functions depend on $L$ [this notation is suppressed in Eqs.~(\ref{eq:LMBW}) - (\ref{eq:cavity})]. By inserting $v$ from Eq.~(\ref{eq:particle_wall}) 
into Eq.~(\ref{eq:profile_equil}) and integrating over $z_1$, we obtain after simplification
\begin{multline}
\frac{d N(L)}{d L}= n(L/2;L) \bigg[1+ \int n(z_1;L) \\
\times h(z_1,L/2,R_{12};L) dz_1 d{\bf R}_{12} \bigg],  
\label{eq:N_deriv}
\end{multline}
where we have used the symmetry with respect to the midplane between the surfaces. This is 
formally a first order differential equation for $N$ as a function of $L$ at constant chemical potential, 
i.e., $dN/dL=f(N,L)$, where $f$ is the right-hand side of Eq.~(\ref{eq:N_deriv}) that implicitly depends 
on $N$. 
In order to solve it we must have a boundary value $N_0=N(L_0)$ for $N$, where $L_0$ is some suitable slit width. This value can be obtained by selecting a large $L_0$ so the density oscillations in the middle of the slit have decayed to a large extent and the density there virtually coincides with the bulk density, i.e., $n(0)=n_0$. Thereby, one solves Eqs.~(\ref{eq:LMBW}) - (\ref{eq:cavity}) for $L=L_0$ by using a value $N=N_0$ such that $n(0)=n_0$ is fulfilled. The value $N_0$  thus obtained can be used as boundary value for the integration of Eq.~(\ref{eq:N_deriv})  to smaller slit widths where the fluid consequently will be in equilibrium with a bulk fluid of density $n_0$. To evaluate $f(N,L)$ one must solve Eqs.~(\ref{eq:LMBW}) - (\ref{eq:cavity}) self-consistently at every step in $L$ during the numerical integration of Eq.~(\ref{eq:N_deriv}) with $N=N(L)$ as  boundary condition for 
Eq.~(\ref{eq:LMBW}). In this manner we obtain the density profiles and pair-distribution functions 
for all wall separations $L\le L_0$.

\subsection{Computational details}
	
In practice, the numerical solution procedure for Eqs.~(\ref{eq:LMBW}) - (\ref{eq:cavity}) and  (\ref{eq:N_deriv}) starts by calculating the density profile and the pair-correlation functions for a wide 
slit of width $L_0$, in our case $L_0 = 15\sigma$. For this surface separation, 
$N=N_0=11.8\sigma^{-2}$ gives the desired value of $n(0)$, i.e., $n_0 = 0.75\sigma^{-3}$. 
The iterational procedure for this initial solution of Eqs.~(\ref{eq:LMBW}) - (\ref{eq:cavity}) is 
started with a constant density profile and pair-correlation functions that are set equal to zero. 
About five CPU-hours are needed for convergence using eight nodes with a clock frequency of 2.6 GHz on a parallel machine. The cut-off radius is $6.4\sigma$ in $r$-space and $98\sigma^{-1}$ in $k$-space; 200 grid point are applied which give a step $\Delta r$ and $\Delta k$ of about $0.032\sigma$ and $0.49\sigma^{-1}$, respectively. In the $z$-direction the step length $\Delta z$ of the grid is equal to $0.025\sigma$. 
	
Once the density profile and the pair-correlation functions have been obtained, 
Eq.~(\ref{eq:N_deriv}) is used to obtain a new value of $N$ for $L=L_0-\Delta L$, where $\Delta L = 2\Delta z = 0.05\sigma$. Thereby, one can utilize a numerical method such as, for example, 
the Runge-Kutta method, which we have chosen here. Equations~(\ref{eq:LMBW}) - (\ref{eq:cavity}) 
are solved again for this slit width and the new value of $N$.  Then the whole procedure is repeated 
for $L-\Delta L$ etc. The start values for the density profile at each new slit width are obtained from 
Eq.~(\ref{eq:profile_equil}), which gives a rather accurate new profile, and the old pair-correlation functions are used as start values for the new ones. Thanks to these good start values in the iterations the convergence becomes very fast, so only another five CPU-hours are needed to obtain self-consistent pair-correlation functions and density profiles for all the slit widths at interval $\Delta L$ down to one hard sphere diameter. Note that the algorithm parallelizes very well so it can 
be even faster if more cores are used in the computer.

\section{Results and discussion}

\subsection{Anisotropic structure factor}

In a physically appealing picture, the experimentally accessible anisotropic structure factor 
$S({\bf q})$ of Refs.~\onlinecite{nygard09a,nygard12} is given by [see the Appendix for details, 
Eqs.~(\ref{eq:S_q_2}) and (\ref{eq:S_q_aver})] 
\begin{equation}
S({\bf q}) = 1+ \int \langle n({\bf r}) h({\bf r},{\bf 0})\rangle
\mathrm{e}^{i{\bf q}\cdot {\bf r}} d{\bf r},  
\label{eq:S_q}
\end{equation}
with {\bf q} denoting the scattering vector. The coordinate system is here placed with the origin 
at the center of a particle, coordinate ${\bf 0}$,  and follows the particle during its motion. 
The vector ${\bf r}$ is a 
position vector that starts from the particle center. The brackets depict an average over all particles 
in the slit, i.e., $\langle n({\bf r}) h({\bf r},{\bf 0})\rangle$ denotes the correlation function for the density distribution around a particle, averaged over all particle positions and weighted with the probability of finding each particle there [cf. Eq.~(\ref{eq:local})]. In other words, the anisotropic $S({\bf q})$ 
probes in a direct manner the 
ensemble-averaged local density correlations in the confined fluid. In the rest of the paper we will denote $\langle n({\bf r}) h({\bf r},{\bf 0})\rangle$ as the averaged local density correlation function.  

\begin{figure}
\centering\includegraphics[width=8.5cm]{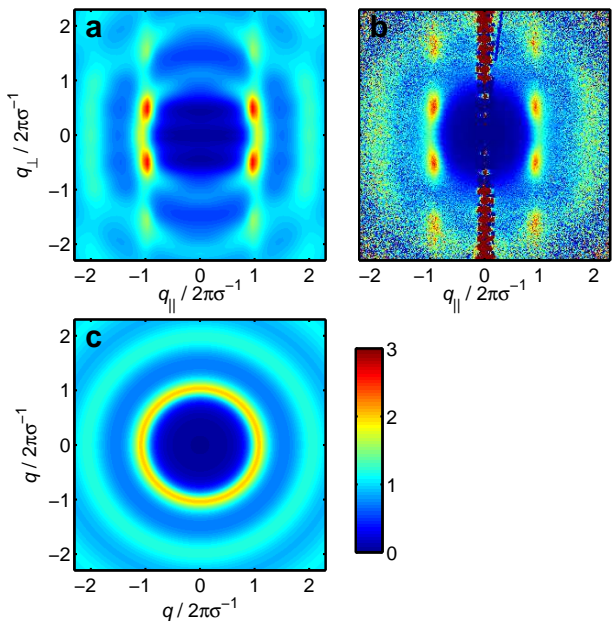}
\caption{Anisotropic structure factor for a hard-sphere fluid confined between hard 
planar surfaces. 
(a) Theoretical and (b) experimental $S(q_{\perp},q_{\parallel})$ are shown for a reduced slit width 
of $L = 2.10\sigma$ and bulk number density $n_0 = 0.75\sigma^{-3}$. The dark red feature at 
$q_{\parallel} = 0$ in the experimental data is diffraction from the confining channel array, which 
should be neglected in the comparison. The experimental data are taken from 
Ref.~\onlinecite{nygard12}.
(c) The corresponding isotropic bulk $S(q)$ for $n_0 = 0.75\sigma^{-3}$. 
\label{fig:exp}
} 
\end{figure}

The visualization of $S({\bf q})$ warrants a brief comment. Since the system has planar symmetry, 
we can without loss of generality write the density profile as $n({\bf r})=n(z)$ and the total 
pair-correlation function as $h({\bf r},{\bf 0})=h(z,R,{\bf 0})$. Here, the $z$ axis is perpendicular 
to the surfaces, $R=|{\bf R}|$, and ${\bf R}=(x,y)$ is directed parallel to the surfaces, i.e., 
{\bf R} is the in-plane component of {\bf r}. Therefore, Eq.~(\ref{eq:S_q}) simplifies to 
\begin{equation}
S(q_{\perp},q_{\parallel}) = 1+ \int \langle n(z) h(z,R,{\bf 0})\rangle
\mathrm{e}^{i (q_{\perp}z+{\bf q}_{\parallel}\cdot {\bf R})} dz d{\bf R},  
\label{eq:S_q_components}
\end{equation}
where $q_{\perp}$ and ${\bf q}_{\parallel}$ denote the out-of-plane and in-plane components 
of the scattering vector, respectively, and $q_{\parallel}=|{\bf q}_{\parallel}|$. Throughout this 
paper, we plot for clarity also negative values of $R$. In these plots, $R$ should be interpreted 
as a coordinate along a straight line in the $xy$ plane through the origin.

Recently, we demonstrated a remarkable agreement between experimental and theoretical 
anisotropic structure factors $S(q_{\perp},q_{\parallel})$ for a hard-sphere fluid confined between 
hard planar surfaces.\cite{nygard12} The quantitative 
agreement is for convenience exemplified in Fig.~\ref{fig:exp} for a reduced slit width of 
$L = 2.10\sigma$. Here, and throughout this study, the confined fluid is kept in equilibrium with a 
bulk fluid reservoir, with the bulk number density $n_0 = 0.75\sigma^{-3}$. 
For comparison, we also present the corresponding bulk structure factor $S(q)$, 
where $q = \vert {\bf q} \vert$. The latter is obtained by solving the isotropic Ornstein-Zernike equation 
within the PY approximation, i.e., the isotropic counterparts of Eqs.~(\ref{eq:OZ}), (\ref{eq:PY}), 
and (\ref{eq:cavity}), and using the hard particle-particle interaction potential in 
Eq.~ (\ref{eq:particle}).\cite{hansen06}   
In contrast to the bulk $S(q)$, both theoretical and experimental $S(q_{\perp},q_{\parallel})$ 
exhibit anisotropy, most strongly manifested as distinct peaks around
$(q_{\perp},q_{\parallel}) \sim (\pm 1/2, \pm 1)$ (in units of $2\pi\sigma^{-1}$) and lobes 
at larger scattering vectors. The excellent agreement between theoretical and experimental 
$S(q_{\perp},q_{\parallel})$ as shown here (and for several slit widths in Ref.~\onlinecite{nygard12})
evidence the accuracy, at the pair-distribution level, of the adopted theoretical scheme. 

\begin{figure}
\centering\includegraphics[width=8.5cm]{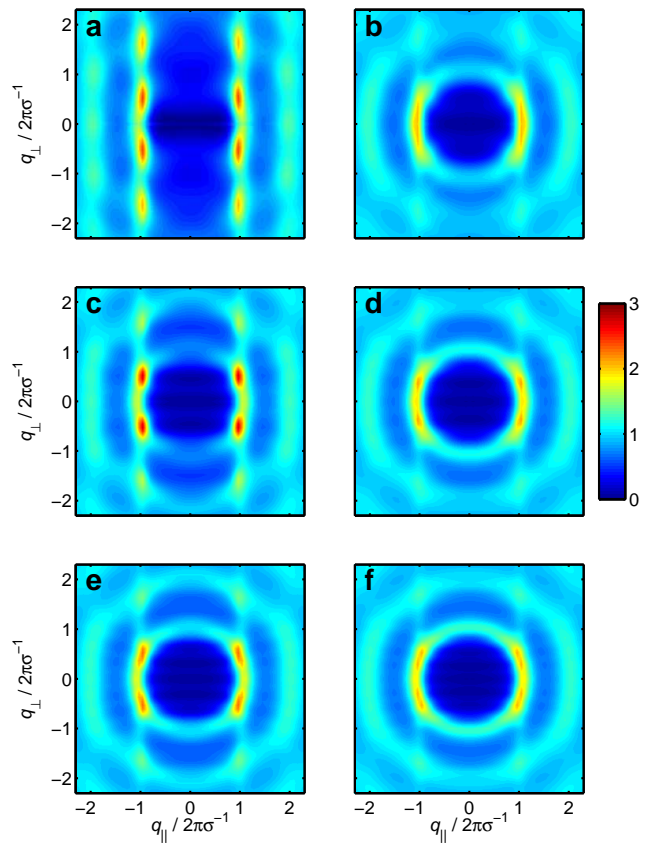}
\caption{Theoretical anisotropic structure factor as in Fig.~\ref{fig:exp}, but for different slit 
widths. The reduced slit widths are (a) $L = 1.05\sigma$, (b) $1.60\sigma$, (c) $2.05\sigma$, 
(d) $2.55\sigma$, (e) $3.00\sigma$, and (f) $3.50\sigma$.
\label{fig:S_q}
} 
\end{figure}

In order to gain further insight into the slit-width dependence of the ensemble-averaged local 
density correlation function in the confined fluid, we present in Fig.~\ref{fig:S_q} the theoretical 
structure factor for a broad range of confining slit widths. Upon increasing the slit width, we observe an 
intriguing sequence of appearances and disappearances of distinct peaks in 
$S(q_{\perp},q_{\parallel})$ [see Video 1 in the supplementary material for a larger set of 
$S(q_{\perp},q_{\parallel})$ plots~\cite{note_suppl}]. Since $S(q_{\perp},q_{\parallel})$ is given 
by the Fourier transform of the ensemble-averaged local density correlation function 
$\langle n(z) h(z,R,{\bf 0})\rangle$ according to Eq.~(\ref{eq:S_q_components}), this 
observation directly implies an alternating sequence of local structural ordering-disordering 
with increasing slit width. 

\begin{figure}
\centering\includegraphics[width=7.5cm]{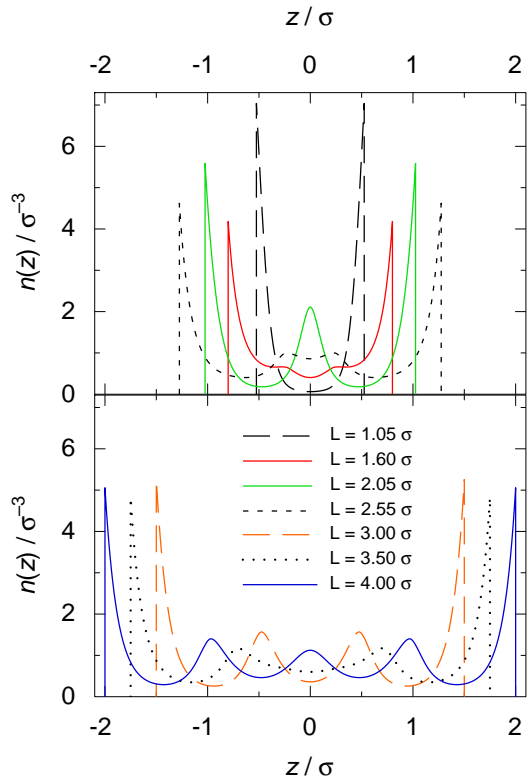}
\caption{Number density profiles $n(z)$ for the hard-sphere fluid confined between 
hard planar surfaces. The reduced slit widths range from $L = 1.05\sigma$ to $4.00\sigma$.
\label{fig:n_z}
} 
\end{figure}

\begin{figure}
\centering\includegraphics[width=7.5cm]{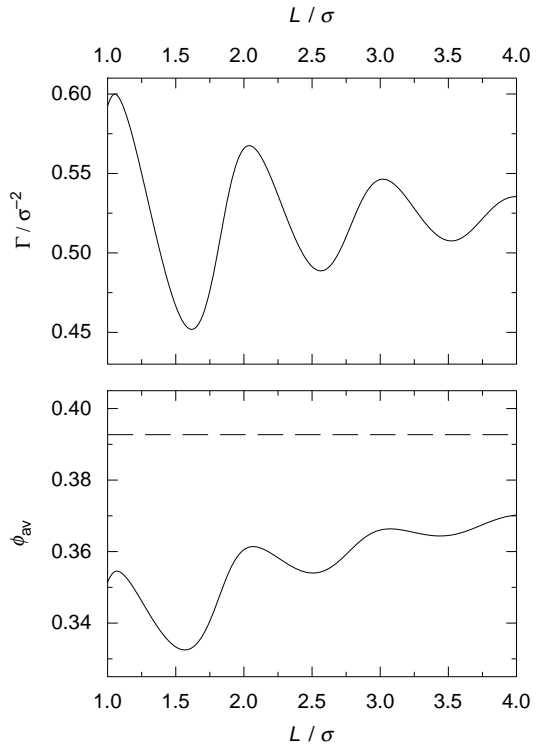}
\caption{Excess adsorption $\Gamma$ and average volume fraction 
$\phi_{av}$ of hard spheres in the slit between two surfaces 
as functions of surface separation. The dashed curve shows the volume fraction in bulk.
\label{fig:gamma_phi}
} 
\end{figure}

We emphasize that the sequence of local ordering-disordering phenomena of Fig.~\ref{fig:S_q} is 
not observable in the traditionally studied density profiles of confined fluids. This is exemplified 
in Fig.~\ref{fig:n_z}, which presents the number density profile $n(z)$ for various slit widths. 
The layered structure between the walls is developed maximally 
for surface separations that are close to an integer multiple of the sphere diameter, while 
for intermediate surface separations the layering is less well developed. We note that the 
two shoulders in the profile for $L=1.60\sigma$ (red curve) have also been found in grand 
canonical simulations.\cite{mittal07} The adsorption excess of particles between the surfaces, 
defined as $\Gamma = \int_{-L/2}^{L/2}[n(z)-n_0]dz$ and displayed in Fig.~\ref{fig:gamma_phi}, 
has peaks at the separations with maximal layering and troughs when the layering is weak. 
The average volume fraction of hard spheres in the slit, 
$\phi_{av}= (\pi\sigma^3/6H) \int_{-L/2}^{L/2}n(z)dz$, 
also presented in Fig.~\ref{fig:gamma_phi}, shows a similar pattern. The chosen slit widths in 
Fig.~\ref{fig:S_q} coincide approximately with subsequent maxima [Figs.~\ref{fig:S_q}(a), 
\ref{fig:S_q}(c), and \ref{fig:S_q}(e)] and minima [Figs.~\ref{fig:S_q}(b), \ref{fig:S_q}(d), and 
\ref{fig:S_q}(f)] in the adsorption excess $\Gamma$.

Importantly, while the peaks in $n(z)$ are more diffuse for slit widths close to minima compared 
to maxima in $\Gamma$, the density profiles of Fig.~\ref{fig:n_z} do not exhibit any qualitative 
changes with increasing slit width which could be interpreted as signatures of  local 
ordering-disordering phenomena at the pair-distribution level. Clearly, much can still be learned 
about confined fluids, even the extensively studied hard-sphere fluid between hard planar 
surfaces, by probing the system at the pair-distribution level. 

It should be noted that $S(q_{\perp},q_{\parallel})$ of the disordered fluid is qualitatively different 
during the transition from $2 \rightarrow 3$ particle layers [Fig.~\ref{fig:S_q}(b)] compared 
to the subsequent transitions from $3 \rightarrow 4$ [Fig.~\ref{fig:S_q}(d)] and $4 \rightarrow 5$ 
particle layers [Fig.~\ref{fig:S_q}(f)]. In particular, the former $S(q_{\perp},q_{\parallel})$ 
exhibits maxima at $(q_{\perp}, q_{\parallel}) \sim (0,\pm 2\pi\sigma^{-1})$, in contrast 
to the latter two cases, indicating a qualitative change in the packing frustration 
of particles for $L \approx 2\sigma$. Indeed, a careful inspection of the number density 
profiles of Fig.~\ref{fig:n_z} (see Video 2 in the supplementary material for a larger set of slit 
widths\cite{note_suppl}) 
verifies this assertion. The transition from $2 \rightarrow 3$ particle layers is found to 
proceed via buckling of the particle layers next to the solid surfaces, akin to so-called 
buckling transitions in thin crystalline films\cite{neser97,schmidt97,fortini06} (for a review 
on buckling transitions, we refer the reader to Ref.~\onlinecite{lowen10}). In contrast, the new particle 
layers are formed in the center of the slit during subsequent layering transitions. It should be 
noted, however, that the peaks at $(q_{\perp}, q_{\parallel}) \sim (0,\pm 2\pi\sigma^{-1})$ 
in $S(q_{\perp},q_{\parallel})$ are observable only in a very narrow range of slit widths, 
$L \approx 1.60\sigma - 1.70\sigma$. Moreover, minor deviations from the ideal system studied 
here, such as size polydispersity of particles and not perfectly parallel planar walls
in the experimental system, may preclude observation of this subtle packing effect. 
Consequently, the peaks at $(q_{\perp}, q_{\parallel}) \sim (0,\pm 2\pi\sigma^{-1})$  
in $S(q_{\perp},q_{\parallel})$  were not experimentally observed in Ref.~\onlinecite{nygard12}.

\begin{figure}
\centering\includegraphics[width=8.5cm]{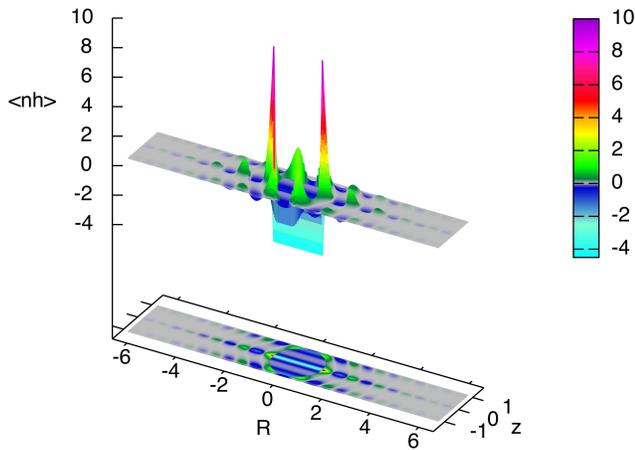}
\caption{Ensemble-averaged local density correlation function 
$\langle n(z) h(z,R,{\bf 0})\rangle$ 
for the reduced slit width $L=1.05\sigma$. In the bottom part of the figure a contour plot of the function is shown and in the top part the same plot is illustrated in a 3D manner with peak heights proportional to the function value. The gray color denotes a narrow interval around the value zero.}
\label{fig:nh_3D}
\end{figure}

 \subsection{Anisotropic local order}

 In order to obtain a real-space picture of the local structural order, we present in 
Fig.~\ref{fig:nh_3D}  the ensemble-averaged local density correlation function 
$\langle n(z) h(z,R,{\bf 0})\rangle$ for the reduced slit width of $L=1.05\sigma$. 
The differences compared to the bulk counterpart $n_0h(r)$ are striking.  
First, the packing of particles leads to highly anisotropic, periodically modulated 
local density correlations, in stark contrast to the isotropic bulk counterpart. 
Second, the peaks in $\langle n(z) h(z,R,{\bf 0})\rangle$ are significantly more pronounced 
compared to the bulk $n_0h(r)$, indicative of the enhanced local order in the former case.  
Third, $\langle n(z) h(z,R,{\bf 0})\rangle$ exhibits structure inside the excluded volume around 
position {\bf 0}, in contrast to bulk fluids. This latter phenomenon can be understood as follows. 
The pair-distribution function $g(z,R,{\bf 0})$ vanishes within the excluded volume around 
{\bf 0}. Consequently, the change in local order, relative to the average structure $n(z)$, 
becomes $\langle n(z) g(z,R,{\bf 0}) - n(z) \rangle = \langle n(z) h(z,R,{\bf 0})\rangle$, 
which equals $-\langle n(z) \rangle$ for 
$z^2+R^2 < \sigma^2$. For bulk fluids $\langle n(z) \rangle$ is simply a constant, whereas for 
confined fluids it is in essence an autocorrelation of $n(z)$, leading to the negative and 
$R$-independent periodic structure inside the excluded volume in the latter case.  
We emphasize that the complexity of $\langle n(z) h(z,R,{\bf 0})\rangle$, as presented here, 
hampers simple analysis of the ensuing $S(q_{\perp},q_{\parallel})$; a proper analysis 
of experimental structure factors from confined fluids, whether colloidal suspension in 
slits\cite{nygard09a,nygard12} or molecular liquids confined in mesostructured porous 
matrices,\cite{bruni98,morineau03} necessitates the calculation of the underlying pair 
distributions theoretically. 

\begin{figure*}
\centering\includegraphics[width=16cm]{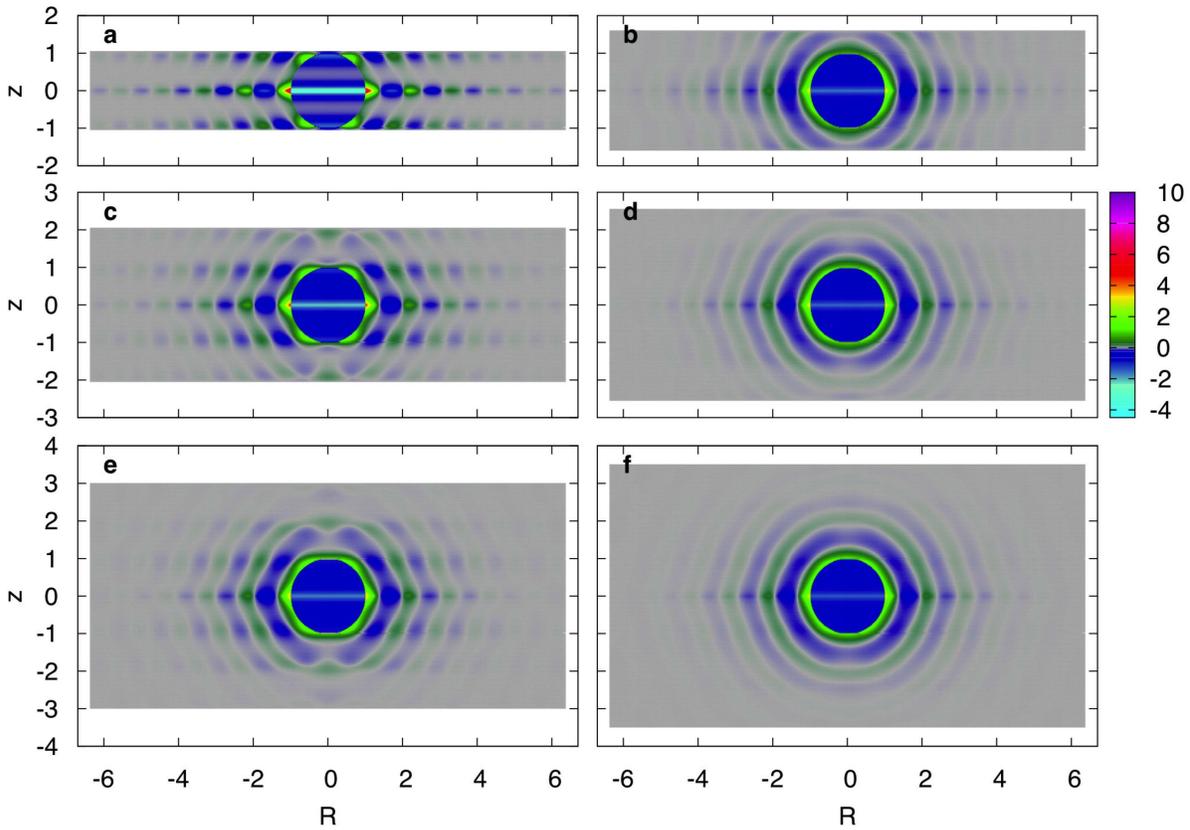}
\caption{Ensemble-averaged local density correlation function 
$\langle n(z) h(z,R,{\bf 0})\rangle$ 
corresponding to the anisotropic structure factors of Fig.~\ref{fig:S_q}. The reduced slit widths 
are (a) $L = 1.05\sigma$, (b) $1.60\sigma$, (c) $2.05\sigma$, (d) $2.55\sigma$, (e) $3.00\sigma$, 
and (f) $3.50\sigma$. 
\label{fig:nh}
} 
\end{figure*}

Now we are in a position to analyze the slit-width dependence of the structure factor. 
In Fig.~\ref{fig:nh}, we present the ensemble-averaged local density correlation function 
$\langle n(z) h(z,R,{\bf 0})\rangle$ for the slit widths of Fig.~\ref{fig:S_q} (see Videos 3 and 4 
in the supplementary material for a larger set of $\langle n(z) h(z,R,{\bf 0})\rangle$ 
plots~\cite{note_suppl}). The complex pattern 
around the central particle at the origin arises from a compromise between a planar layering of 
particles between the surfaces and a spherical layering induced by the particle. As anticipated 
based on $S(q_{\perp},q_{\parallel})$, we observe a sequence of local ordering in 
$\langle n(z) h(z,R,{\bf 0})\rangle$, with the fluid alternating between a periodic density pattern 
when the central particle penetrates layers that are in a quite ordered state 
and a more isotropic, bulk-like density pattern when the layers are in a more disordered 
(i.e., frustrated) state. 

We note that the strong anisotropy observed for surface separations close to an integer multiple 
of the particle diameter [Figs.~\ref{fig:nh}(a), \ref{fig:nh}(c), and \ref{fig:nh}(e)] is much less 
developed for 
$L> 3.0\sigma$. The plot for $L= 4.0\sigma$ (see Videos 3 and 4 in the supplementary 
material~\cite{note_suppl}) shows only slightly more structure than that for $L= 3.5\sigma$, 
Fig.~\ref{fig:nh}(f). Similarly, the corresponding $S(q_{\perp},q_{\parallel})$ plots in 
Fig.~\ref{fig:S_q} (and in Video 1~\cite{note_suppl} in the supplementary material) become 
more bulk-like for $L> 3.0\sigma$ [cf. Fig.~\ref{fig:exp}(c)], with the distinct peaks around 
$(q_{\perp},q_{\parallel}) \sim (\pm \pi\sigma^{-1}, \pm 2\pi\sigma^{-1})$ becoming 
strongly suppressed and the lobes at larger scattering vectors becoming nearly isotropic. 
Intriguingly, recent theoretical work on the diffusivity in confined hard-sphere fluids has 
revealed a similar slit-width dependence, with the oscillatory behavior of the diffusion coefficients 
as a function of slit width being strongly suppressed for $L>3\sigma$.\cite{mittal08,lang10}  
On a microscopic level, the diffusivity depends on the local density of the confined fluid; 
more ordered fluids have a larger free volume and hence a larger diffusivity.\cite{mittal08} 
However, more theoretical work is needed in order to formally connect the 
$\langle n(z) h(z,R,{\bf 0})\rangle$'s of Fig.~\ref{fig:nh} to the findings of 
Refs.~\onlinecite{mittal08,lang10}. 

We also observe a subtle, yet important, difference in Fig.~\ref{fig:nh} between the local density 
correlation $\langle n(z) h(z,R,{\bf 0})\rangle$ of the disordered fluids, which leads to the qualitatively 
different behavior of the peaks in $S(q_{\perp},q_{\parallel})$ as discussed above, namely, the 
peaks located at $(q_{\perp}, q_{\parallel}) \sim (0,\pm 2\pi\sigma^{-1})$ for separations 
$L \approx 1.60\sigma - 1.70\sigma$, which are split into two peaks each with nonzero 
$q_{\perp}$ for other surface separations. In Figs.~\ref{fig:nh}(d) and \ref{fig:nh}(f), $L=2.55\sigma$ 
and $3.50\sigma$, we observe spatial correlations between particles in neighboring 
layers, similar to the sixfold correlations observed for  $L=1.05\sigma$,  $2.05\sigma$, 
and $3.00\sigma$ [Figs.~\ref{fig:nh}(a), \ref{fig:nh}(c), and \ref{fig:nh}(d)], but less pronounced 
(keep in mind that the present system exhibits planar symmetry). 
For $L=1.60\sigma$ [Fig.~\ref{fig:nh}(b)], on the other hand, the peaks are smeared out 
in the $z$ direction, which leads to the peaks in $S(q_{\perp},q_{\parallel})$ at zero $q_{\perp}$.

\begin{figure}
\centering\includegraphics[width=7.5cm]{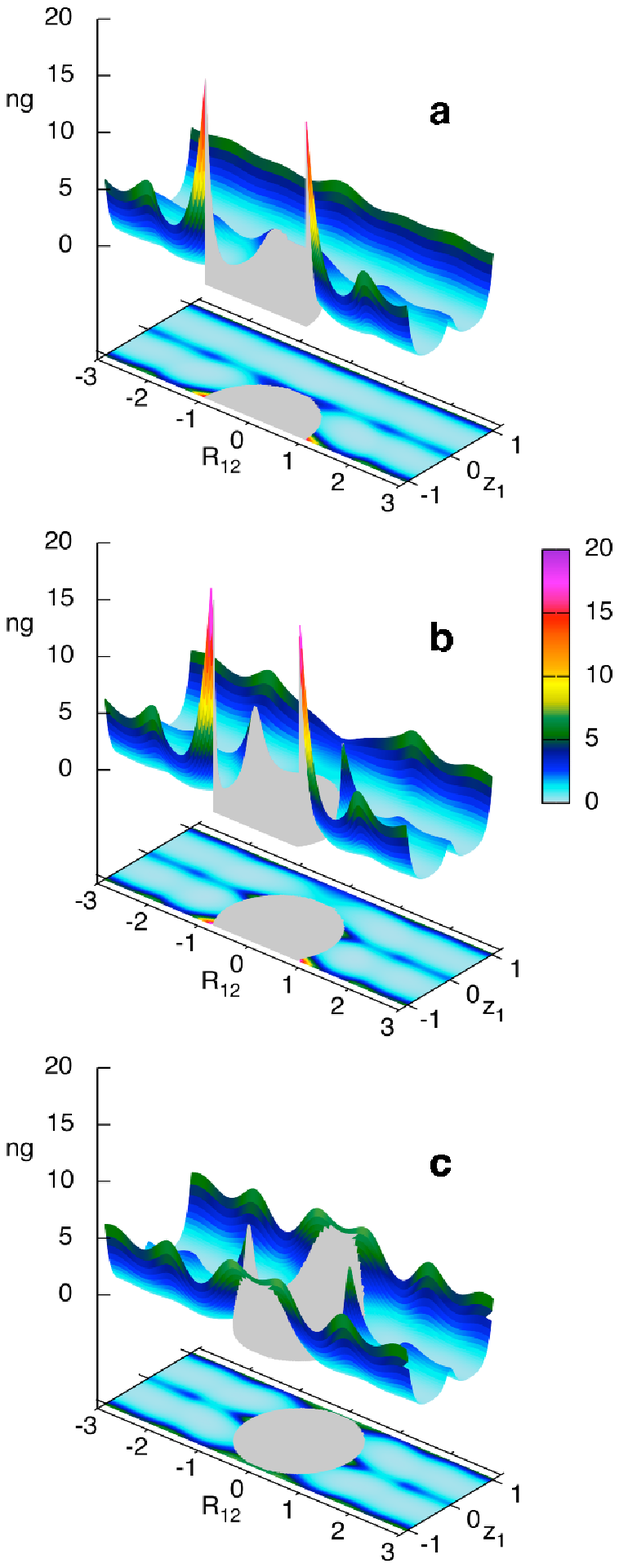}
\caption{Local density $n(z_1)g(z_1,z_2,R_{12})$ at coordinate $({\bf R}_{12},z_1)$ 
around a particle in the slit between two hard surfaces, when the particle is located on the $z$ axis 
at coordinate $z_2$. Data are shown for the reduced slit width $L = 2.05\sigma$ and three 
different particle positions: (a) in contact with one surface, (b) at the density minimum, and (c) in 
the slit center. The gray region depicts the excluded volume around the particle. 
\label{fig:ng}
} 
\end{figure}

\subsection{Anisotropic local density}

The local correlation function $\langle n(z) h(z,R,{\bf 0})\rangle$, which is an ensemble average 
over all particles in the slit, can be decomposed into underlying local densities 
$n({\bf r}_1)g({\bf r}_1,{\bf r}_2)$ for various particle positions ${\bf r}_2$ [see the Appendix and Eq.~(\ref{eq:local}) below for the relationship between these kinds of entities]. 
To facilitate the understanding of the meaning of $\langle nh \rangle$ we here present 
the function $ng$ for a few cases (in Ref.~\onlinecite{kjellander91} some other plots of this 
function can be found for a somewhat lower bulk density).
Due to planar symmetry, we have $n({\bf r}_1)g({\bf r}_1,{\bf r}_2) \equiv n(z_1)g(z_1,z_2,R_{12})$ 
where $R_{12}=|{\bf R}_{12}|$ and ${\bf R}_{12}=(x_1-x_2,y_1-y_2)$, so $R_{12}$ denotes 
the in-plane projection of $\vert {\bf r}_1-{\bf r}_2\vert$. In graphical representations of this function, 
it is convenient to let the $z$ axis go through the particle center, i.e., we select  ${\bf r}_2 = (0,0,z_2)$. 
Then  the function $n(z_1)g(z_1,z_2,R_{12})$ states the density at a point 
${\bf r}_1 =({\bf R}_{12},z_1) = (x_1,y_1,z_1)$, when a particle is located at $(0,0,z_2)$. Again, 
we plot for clarity also negative values of $R_{12}$, i.e., in the following plots $R_{12}$ is to be 
interpreted as a coordinate along a straight line in the $xy$ plane through the origin.  

In Fig.~\ref{fig:ng}, we show examples of the local density $n(z_1)g(z_1,z_2,R_{12})$ 
for the reduced slit width $L = 2.05\sigma$. In these plots, the particle is positioned 
(a) in contact with one surface at $z_2 = -1.025\sigma$, (b) at the density minimum at 
$z_2 = -0.475\sigma$, and (c) in the slit center at $z_2=0$ (a density maximum). 
For more particle positions $z_2$, we refer to Video~5 in the supplementary material.\cite{note_suppl}  
The anisotropy in local density depends strongly on the particle position $z_2$, or 
more specifically on the packing conditions for other particles given a particle at $z_2$. 
Most notably, in Figs.~\ref{fig:ng}(a) and \ref{fig:ng}(b) the particle density in the wedge-like 
section formed between the particle and the nearby wall is strongly enhanced, resulting in a 
local number density of up to $18.0\sigma^{-3}$ and $22.9\sigma^{-3}$, respectively.  
The excluded volumes of the particle and the wall meet there and form a section where 
other particles can come in but not not pass through. Particles will remain there for a relatively 
long time because when they try to escape they will usually be pushed in again by collisions with 
the surrounding particles.  
We note that similar local density enhancements have also been observed in binary 
hard-sphere mixtures and discussed in terms of depletion interactions.\cite{botan09} 
The enhancement in the local density $n(z_1)g(z_1,z_2,R_{12})$ relative to the singlet  
density $n(z_1)$ is given by the pair-distribution function $g(z_1,z_2,R_{12})$. In the inner 
part of the wedge-like section for these two cases $g$ reaches $3.2$ and $4.1$, respectively. 
For comparison, the maximum value of the local density in Fig.~\ref{fig:ng}(c), where no such 
wedge-like sections induced by overlapping excluded volumes exist, is a factor of $\sim 3$ 
smaller compared to those of Figs.~\ref{fig:ng}(a) and \ref{fig:ng}(b). 

The consequences of the penetration of the central particle into the particle layers between the 
walls as seen in Fig.~\ref{fig:ng} are also apparent in $\langle n(z) h(z,R,{\bf 0})\rangle$ of 
Fig.~\ref{fig:nh}(c). In particular, 
the six-fold correlations mentioned above are readily observed when the particle position $z_2$ 
is close to a density maximum like in Fig.~\ref{fig:ng}(c). Note that the large peaks in 
$\langle n(z) h(z,R,{\bf 0})\rangle$ for $z=0$ originate from $n(z_1)g(z_1,z_2,R_{12})$ with 
$z_1=z_2$ for various particle positions $z_2$. A major contribution to these peaks comes from the 
case with the particle in contact with a wall, Fig.~\ref{fig:ng}(a), i.e., from the main density peaks 
at the wall which we discussed above. A substantial contribution also comes from cases like $z_2=0$, 
Fig.~\ref{fig:ng}(c); in this case from the density peaks at $z_1=0$ .

In the general case, a detailed quantitative analysis of local order in confined fluids based 
on $n({\bf r}_1)g({\bf r}_1,{\bf r}_2)$ may be a huge task, simply because of the large number 
of independent variables, ${\bf r}_1$ and ${\bf r}_2$. As we have seen, there is a strong variation 
in both (i) $n({\bf r}_1)g({\bf r}_1,{\bf r}_2)$ with ${\bf r}_1$ and ${\bf r}_2$ and 
(ii) the probability of finding a particle 
at ${\bf r}_2$ [which is proportional to $n({\bf r}_2)$]. This applies even when these functions 
can be calculated without undue effort. Analysis of $\langle n({\bf r}) h({\bf r},{\bf 0})\rangle$ 
is in this respect more convenient. Nevertheless, the explicit analysis of 
pair distributions is important for determining other properties of the system, such as the net 
force acting on a particle at various positions in the confined space.\cite{kjellander91}

\section{Conclusion}

The ensemble-averaged local density correlation function $\langle n(z) h(z,R,{\bf 0})\rangle$, 
as introduced and studied in this paper, exhibits two notable advantages. 
First, the local density $n(z_1)g(z_1,z_2,R_{12})$ is a multidimensional quantity, which 
depends on two positions relative to the confining surfaces; one of the positions is occupied 
by a particle and the other gives the position where the particle density is measured. In contrast, 
$\langle n(z) h(z,R,{\bf 0})\rangle$ is an ensemble average over all particles in the 
slit, which means that unlikely particle configurations are effectively filtered out. In essence, 
we have coarse-grained out one spatial dimension in $\langle n(z) h(z,R,{\bf 0})\rangle$ 
compared to $n(z_1)g(z_1,z_2,R_{12})$, which greatly facilitates the analysis of pair distributions 
in confined fluids. Second, $\langle n(z) h(z,R,{\bf 0})\rangle$ is directly accessible in 
x-ray scattering experiments, thereby allowing a quantitative comparison between experiment 
and theory in terms of the microscopic structure at the pair-distribution level. We foresee 
extensive studies of $\langle n(z) h(z,R,{\bf 0})\rangle$ for different particle-wall and 
particle-particle interaction potentials.  

This study focuses on a simple model system -- the extensively studied hard-sphere 
fluid confined between smooth and hard planar surfaces. However, the main finding 
reported here, a packing-frustration-induced alternating sequence of a strongly anisotropic, 
periodically modulated versus a more isotropic local order, is expected to be a general 
phenomenon. As mentioned in 
the Introduction, the hard-sphere fluid confined between hard surfaces can be regarded as a 
good approximation for entropy-dominated fluids: First, the pair distributions of simple dense 
fluids exhibiting short-ranged particle-particle interactions are dominated by the excluded volume 
of the core region, which is contained in the present model. Second, short-ranged particle-wall 
interactions are dominated by the excluded volume at the interface, which is again included in 
the model. Further support for the generality of the observed ordering phenomenon is given by 
the following two examples: (i) anisotropic local densities $n(z_1)g(z_1,z_2,R_{12})$ resembling 
those presented in Fig.~\ref{fig:ng} have previously been reported for confined Lennard-Jones 
fluids\cite{kjellander91b} and (ii) signatures in $S(q_{\perp}, q_{\parallel})$ of local ordering, 
similar to those presented in Fig.~\ref{fig:S_q}, have been experimentally observed in a system 
of charged colloidal particles confined between charged surfaces.\cite{nygard09a}  
We therefore expect the local ordering-disordering phenomenon as observed here to be 
an intrinsic property of a large class of dense simple fluids under spatial confinement.

Finally, we return to the computational effort alluded to in the Introduction. In this work, we have 
determined the anisotropic pair-distribution functions and the structure factor by application of 
integral-equation theory.  In principle, these functions could also be evaluated directly from particle 
configurations obtained by grand-canonical Monte Carlo\cite{mittal07} or molecular 
dynamics\cite{gao97} simulations. However, even with the computing power presently 
available, one would need impracticably long simulations in order to obtain a  reasonable 
statistical accuracy for the entire $n({\bf r}_1)g({\bf r}_1,{\bf r}_2)$. Alternatively, one can determine 
the pair distributions point-wise in simulations using the Widom insertion method, provided the 
fluid is not so dense that this method becomes too inefficient. Such a Monte Carlo approach has 
previously been compared with an integral-equation theory, similar to the one used here, in the 
case of inhomogeneous electrolytes;\cite{greberg96} for the corresponding amount of 
pair-distribution data of essentially equal accuracy, the integral-equation approach was 
found to be many thousands  times more efficient in CPU time than the simulations. 

For the present work, the computations of all results presented were carried out in less than 
10 hours of CPU time (see Sec.~II for details). This included calculations of pair 
correlations and density profiles  for all surface separations from  $L=15.00\sigma$ to $1.00  \sigma$ 
with a resolution $\Delta L=0.05  \sigma$. Only a small fraction of these results is presented here. 
Hence, we see no genuine computational barriers precluding studies of confined fluids, or more 
generally inhomogeneous fluids, at the pair-distribution level. Similar conclusions were drawn in 
a recent computational study of electrolytes confined between two dielectric planar 
surfaces.\cite{zwanikken13}

\begin{acknowledgments}
R.K. and K.N. acknowledge support from the Swedish Research Council 
(Grant Nos. 621-2009-2908 and 621-2012-3897, respectively). 
The computations were supported by the Swedish National Infrastructure for Computing 
(SNIC 001-09-152) via PDC. 
\end{acknowledgments}

\appendix*

\section{Anisotropic structure factor}

The experimentally accessible anisotropic structure factor of Refs.~\onlinecite{nygard09a,nygard12} 
is given by
\begin{equation}
S({\bf q}) = 1+\frac{1}{M}\int \int 
n({\bf r}_1)n({\bf r}_2) h({\bf r}_1,{\bf r}_2)
\mathrm{e}^{i{\bf q}\cdot({\bf r}_1-{\bf r}_2)} d{\bf r}_1d{\bf r}_2,  
\label{eq:S_q_2}
\end{equation}
where $M$ denotes the total number of particles in the confining slit ($M=NA$, where $A$ 
is the area of the wall surface). In a more intuitive picture, 
$S({\bf q})$ probes the local density correlation around a particle, averaged over all particles in the 
slit. Formally, this is obtained by averaging over all possible positions ${\bf r}_2$ across the 
confining channel, weighted with the probability density $p({\bf r}_2)=n({\bf r}_2)/M$ of finding 
the particle at position ${\bf r}_2$. In practice, this is readily achieved by fixing the coordinate 
system {\bf 0} in Eq.~(\ref{eq:S_q_2}) on the particle at position ${\bf r}_2$, leading to 
\begin{equation}
S({\bf q}) = 1+ \int \langle n({\bf r}) h({\bf r},{\bf 0})\rangle
\mathrm{e}^{i{\bf q}\cdot {\bf r}} d{\bf r},  
\label{eq:S_q_aver}
\end{equation}
where $\langle \cdot \rangle$ is the average with respect to the probability density $p$ 
and ${\bf r}={\bf r}_{12}={\bf r}_1-{\bf r}_2$ is the position vector that starts from the particle center. 
The functions $n$ and $h$ have here been redefined and written with respect to the particle-centered 
coordinate system, i.e., $n({\bf r}+{\bf r}_2)\Rightarrow n({\bf r})$ and 
$h({\bf r}+{\bf r}_2,{\bf r}_2)\Rightarrow h({\bf r},{\bf 0})$. The ensemble-averaged local density 
correlation function can alternatively be written as 
\begin{equation}
\langle n({\bf r}) h({\bf r},{\bf 0})\rangle =  
\int_{-(L+z-\vert z \vert)/2}^{(L-z-\vert z \vert)/2} w(z_2) n(z+z_2) h(z+z_2,z_2,R) dz_2, 
\label{eq:local}
\end{equation}
where the functions in the integrand are written with respect to the coordinate system with origin 
at the midplane of the slit, $w(z_2) = n(z_2) / N$ is the appropriate weight function, 
${\bf r}=({\bf R},z)$, and $R=|{\bf R}|$. 
We note that the available space perpendicular to the confining surfaces equals $2L$ rather 
than $L$, since the largest possible out-of-plane distance between two particles is 
$L$ in both positive and negative directions.


\end{document}